\documentclass[a4paper,10pt]{article}
\usepackage[utf8]{inputenc}
\usepackage{booktabs}
\usepackage{graphicx}
\usepackage{framed}
\usepackage{url}
\usepackage{authblk}

\title{Fruitful symbioses between\\ termites and computers}

\author{Og DeSouza\thanks{\texttt{og.souza@ufv.br}; url: \texttt{http://www.isoptera.ufv.br}}}
\affil{Federal University of Viçosa, Brazil}

\author{Elio Tuci\thanks{\texttt{elt7@aber.ac.uk}}}
\affil{Aberystwyth University, Wales}

\author{Octavio Miramontes\thanks{\texttt{octavio@fisica.unam.mx}}}
\affil{National Autonomous University, Mexico}

\graphicspath{{../../70figs/}}

\date{}
\begin{document}

\maketitle

\begin{abstract}

The living-together of distinct organisms in a single termite nest along with the termite builder colony, is emblematic in its ecological and evolutionary significance. On top of preserving biodiversity, these interspecific and  intraspecific ``symbioses'' provide useful examples of interindividual associations thought to underly transitions in organic evolution. Being interindividual in nature, such processes may involve emergent phenomena and hence call for analytical solutions provided by computing tools and modelling, as opposed to classical biological methods of analysis. Here we provide selected examples of such solutions, showing that termite studies may profit from a symbiotic-like link with computing science to open up wide and new research avenues in ecology and evolution. 

\end{abstract}

\section{Introduction}

Termites are protagonists of processes which are crucial to ecological functioning and whose understanding would unveil puzzles still persistent in evolutionary theory. Specifically, the fact that termites live symbiotically in familiar groups (``colonies'') comprising a highly integrated society, poses them at the upper end of the series of transitions characterizing organic evolution \cite{Maynard-Smith.Szathmary.1997.majortransitionsin} (Table~\ref{tab:transitions}).

Simultaneously, termites often establish symbioses with other organisms (even other termite species) which cohabit termite nests along with their building colony \cite{Florencio.etal.2013.Diet}
. This makes a termite nest a hotspot of diversity, with important consequences for ecosystem functioning \cite{Pringle.etal.2010.Spatial}. It also makes a termite nest a hotspot of evolutionary processes, varying from inter-specific symbiotic relationships to full sociality (this latter being a form of symbiosis occuring at the intraspecific level).

Despite diverse in output, many of these processes are based on intra- or interspecific interactions occurring between individuals which cohabit the same termitarium. Being interindividual in nature, such processes may involve emergent phenomena and hence call for analytical solutions provided by computing modelling, as opposed to classical biological methods of analysis.

Here we provide selected examples on how can computing tools and modelling help better understanding of ecological and evolutionary processes through the study of termite societies and their associated cohabitants. Such examples will involve computer-aided tracking and path analyses, cellular automata simulations, the analyses of emergent properties in networks, and the use of genetic algorithms and artificial life simulations.

\subsection{What are termites?}
Termites are six-legged social animals, majorly tropical, but also present in many temperate countries. They belong to the same order as the coackroaches (Blattodea) and comprise the phylogenetically cohesive infraorder Isoptera \cite{Krishna.etal.2013}. Bees, ants and wasps, which are also social insects, belong to another order, the Hymenoptera. 

Being social, termites live in colonies within nests built by themselves. Such colonies are a familiar group composed by both female and male parents plus their male and female offspring. Having a society composed by male and female members is the biggest difference between termites and other social insects such as ants and bees. In these latter, most members are females (males are short-lived and are produced sazonally). Other differences exist and we shall introduce them along the text, as needed.

\subsection{Why are termites important?}
Termites are quite important for both, practical and theoretical reasons. Although mostly known for their pest status, this is not their sole or their most important trait. Termites are among the rare animals which are able to digest celullosis and this means that they are ultimately responsible for transforming organic carbon into its inorganic form, which is the major source of carbon used by plants. By possessing their own endogenous cellulolitic enzimes in addition to cellulolitic gut symbionts, termites can oxidise nearly 99\% of the carbon they intake \cite{Slaytor2000Energy}, being then among the few animals to take active part in the carbon cycle in earth. Maybe more importantly, they prevent excessive accumulation of dead plant material \cite{DeSouza.etal.2009} and dung from large herbivores \cite{Freymann2008} on the Earth surface . It is worth mentioning that, if we copy termite’s diggestion process we can be more efficient in producing carbon-free fuel because we will be using, \textit{e.g.}, sugar but the whole of the sugarcane to produce ethanol \cite{ScharfBoucias2010Potential}. 

The importance of termites to ecological process on Earth is stressed by their huge biomass: it is estimated that termites along with ants comprise a third of all Amazonia's animal biomass \cite{Fittkau.Klinge.1973}. A similar picture applies to termites in African savanna. Because termites are small their metabolism is faster than, e.g., that of a vertebrate. This would make their astonishing biomass even more significant, as per unit weight and time, termites are prone to process more materials than vertebrates. 

In order to shelter their colonies, termites build nests which may vary from simple excavations in wood or soil to elaborate buildings on the soil surface or on trees. In doing so, termites modify the surrounding environment, allowing the estabilishment of other plant and animal species, either in the vicinity of the nest, on its external walls, or even inside the whole structure. This role as an ``ecosystem engineer" percolates to positively impact the animal and vegetal primary productivity of the region \cite{Pringle.etal.2010.Spatial}. In summary, termites  are central do ecological processes on Earth.

The relevance of termites to theoretical insights into biology stems from the fact that these are the oldest social insect \cite{Grimaldi.Engel.2005.Evolution}, providing therefore, primordial views on the processes ruling the origin and maintenance of sociality. This is quite important, because sociality is considered one of the last transitions in evolution \cite{Maynard-Smith.Szathmary.1997.majortransitionsin}.

Sociality in termites (along with that in all ants, some bees, wasps, and a few other animals) is of a special kind because it can be posed at the highest level of all forms of colonial organisation. In such cases, sterile offspring cohabit with and cooperatively help their parents to raise fertile offspring, and this is called ``eusociality'' \cite{Wilson.1971.InsectSocieties}. This division of reproductive labour estabilishes within the colony distinct groups of individuals 
which share similar behavioural and sometimes morphological traits, the so called ``castes''. Non-reproductive castes are, in termites, comprised by ``workers'' and ``soldiers''. The former are responsible for nest building, maintenance, foraging, etc, while the soldiers, as the name implies, take care of nest defense. In most termites, such sterile castes are irreversible, which means that the members of a termite society are truly dependent on each other for survival and reproductive success. 

\section{The paradox of eusociality}

Eusociality presents an evolutionary paradox \cite{Miramontes.DeSouza.2014.Social}: if adaptive evolution is driven by differential reproduction of individuals, how can sterile members of a social insect colony evolve and persist? Darwin himself commented on these sterile individuals saying that [...] \textit{they represent one special difficulty, which at first appeared to me insuperable, and actually fatal to the whole theory}. As a solution, he proposed that queens which were able to produce cooperative sterile offspring in addition to ordinary fertile ones, would succeed better than those producing only non-cooperative fertile offspring, as these latter queens would not profit from the synergism inherent to cooperative work. 

Nearly a hundred years after, in 1964, an elegant mathematical formalism aimed to solve the riddle was proposed by W. D. Hamilton. It consisted of the so-called ``kin selection'', sometimes also referred to as “inclusive fitness”.  In essence, kin selection predicts that infertile individuals, by cooperating with fertile family members, warrant the transmission of their own genes indirectly, when their kin reproduce. This, of course, will only work if sterile helpers are closely related to their fertile nestmates.

 Haplodiploidy in Hymenoptera (bees, ants, wasps), where males are haploid and females are diploid, seem a key to this puzzle because it fulfills the condition of high relatedness among nestmates. A hymenopteran female can share 75\% of its genes with her nestmate sisters. Haplodiploidy, therefore, warrants higher levels of kinship between females, which, even if abstaining reproduction and helping their mother to raise reproductive sisters, would transfer to the next generation more genes than they could do by direct reproduction. 

 The higher levels of kinship among haplodiploids can be easily spotted by a simple calculation. Consider a fully heterozygous haplodiploid cross:
\begin{center}
    \begin{tabular}{l|r}
        & B\\\hline
      A & AB\\
      a & aB\\\hline
    \end{tabular}
    \end{center}

Because in Hymenoptera diploids are females, all offspring produced from this cross is female (males are produced parthenogenetically). If we then inspect the degree of relatedness between these sisters we will get:
\begin{center}
\begin{tabular}{l|p{0.7cm}p{1cm}}
sisters & AB & aB\\\hline
AB & 1.0 & 0.5\\
aB & 0.5 & 1.0\\\hline
\end{tabular}
\end{center}
That is, in average, sisters are related to each other by: $(1.0+0.5+0.5+1.0)/4=0.75$%

Haplodiploidy, however, is not sufficient to explain the evolution of eusociality. On one hand, there are lots of solitary hymenopteran species which are also haplodiploid, which means that haplodiploidy does not lead necessarily to eusociality. On the other hand, and maybe more importantly, eusociality also occurs among diplo-diploid organisms (those in which both, males and females, are diploids). Among these we find all termite species (\textit{c.a.} 3,000) plus aphids, beetles, shrimps, and naked mole rats -- not to mention humans which may also be said eusocial under certain circumstances. Fully diploid organisms would profit more from their own reproduction than that of their relatives, since they are in average only 50\% akin, as detailed below. In fully heterozygous diplo-diploid cross the following offspring can be produced:
\begin{center}
 \begin{tabular}{l|rp{0.2cm}r}
      & B && b\\\hline
      A & AB && Ab\\
      a & aB && ab \\\hline
    \end{tabular}
    \end{center}

Siblings from such a cross will exhibit the following degree of relatedness:
\begin{center}
\begin{tabular}{l|p{0.7cm}p{0.7cm}p{0.7cm}p{0.7cm}}
siblings & AB  & Ab & aB & ab \\\hline
AB & {1.0} & {0.5} & {0.5} & {0.0} \\
Ab & {0.5} & {1.0} & {0.0} & {0.5} \\
aB & {0.5} & {0.0} & {1.0} & {0.5} \\
ab  & {0.0} & {0.5} & {0.5} & {1.0} \\\hline
\end{tabular}
\end{center}
In such case, therefore, the average relatedness between siblings is:  $$((1*4)+(0.5*8))/16=0.50$$%

The reasons for the existence of steriles among diplo-diploids, therefore, remain unclear. In other words, we are still awaiting for a convincing explanation on why termites are eusocial and at the same time diploid \cite{Thorne.1997}.

\section{A fertile terrain for computing models and tools}
\subsection{Intraspecific interactions and social behaviour}
Termites are, thus, central to the theory of evolution because: (i) they are social and hence are posed at  the pinnacle of evolutionary processes and (ii) their sociality is not easily explained. Why are them social, being not haplo-diploid? 

To answer this question, computing models come in very handy. Being hemimetabolous, termites are highly mobile since birth, unlike bees, ants, or wasps, in which juveniles are grub-like and hence almost sessile. They are also highly interactive and excitable: they exchange food and care among themselves and are able to pass information from one to another member of the colony. 

Termites, hence, seem autonomous interactive mobile agents and as such possess traits highly prone to generate group-level dynamics. An immediate hypothesis arises from this: eusociality can be an emergent phenomenon, arising form interindividual interactions among colony members \cite{Miramontes.DeSouza.2014.Social}. That is, eusociality could stem from  group-level phenomena. If this is so, termites could profit from collective behaviour per se, offsetting small genetic advantages inherent to diplo-diploidy. Sterile castes would, then, profit from cooperation and group living, surviving longer and hence getting more time to pass their genes on indirectly. 

\begin{center}
 \begin{figure}[b]
 \centering
  \includegraphics[scale=0.5]{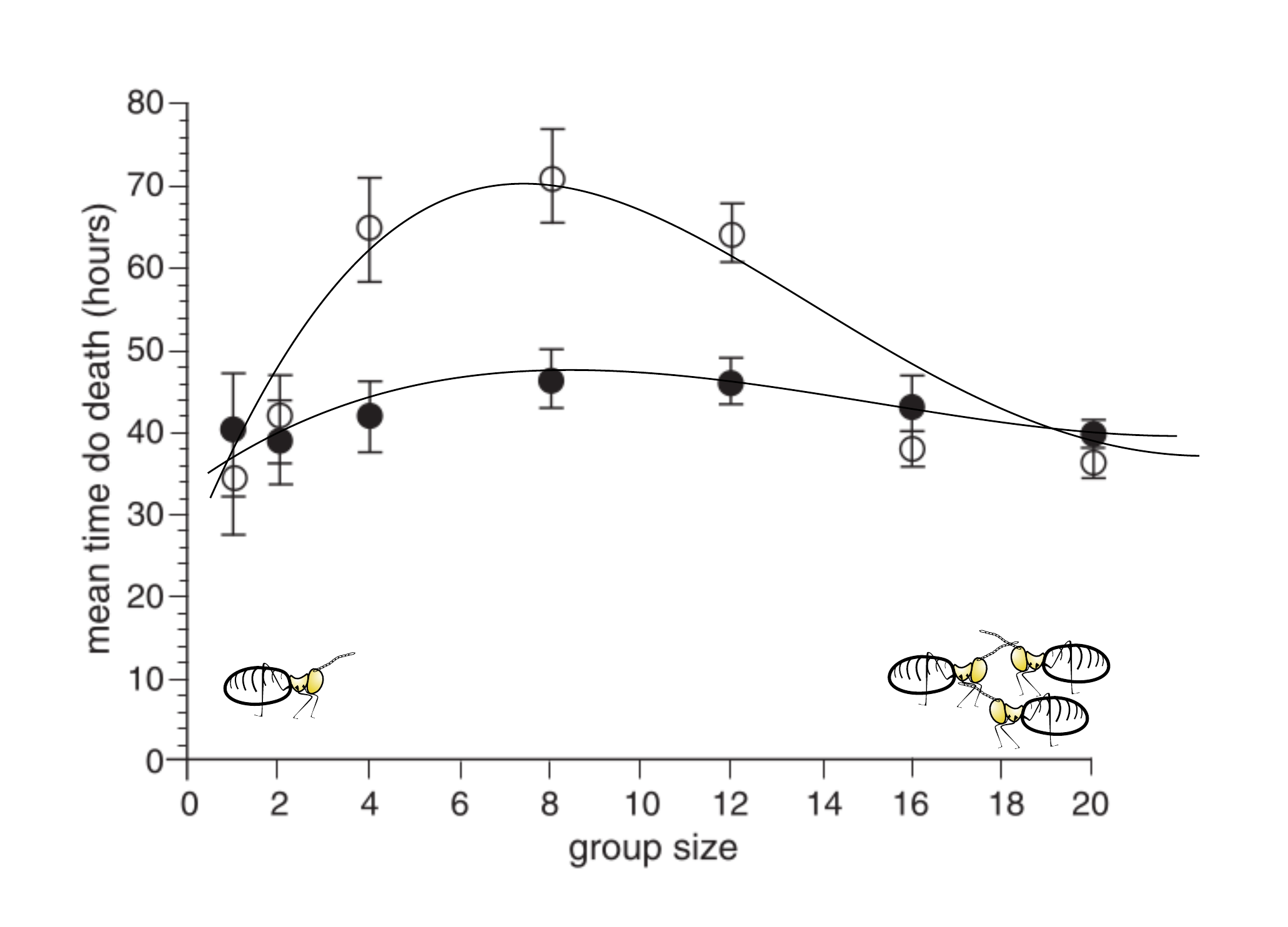}
  \caption{Survival of starving termites as a function of the group size in which they have been confined. The data is split in two categories: termites exposed to insecticides (black dots) and termites which were not exposed to insecticides (hollowed circles), to show that group size effects persist even under strong stress. Survival is measured, from Weibull distribution, as the average number of hours group members spent to die. Group size is the number of individuals confined together. Modified from \cite{DeSouza.etal.2001}. }
  \label{fig:survival-grouped}
 \end{figure}
\end{center}

\begin{center}
 \begin{figure}
 \centering
  \includegraphics[scale=0.6]{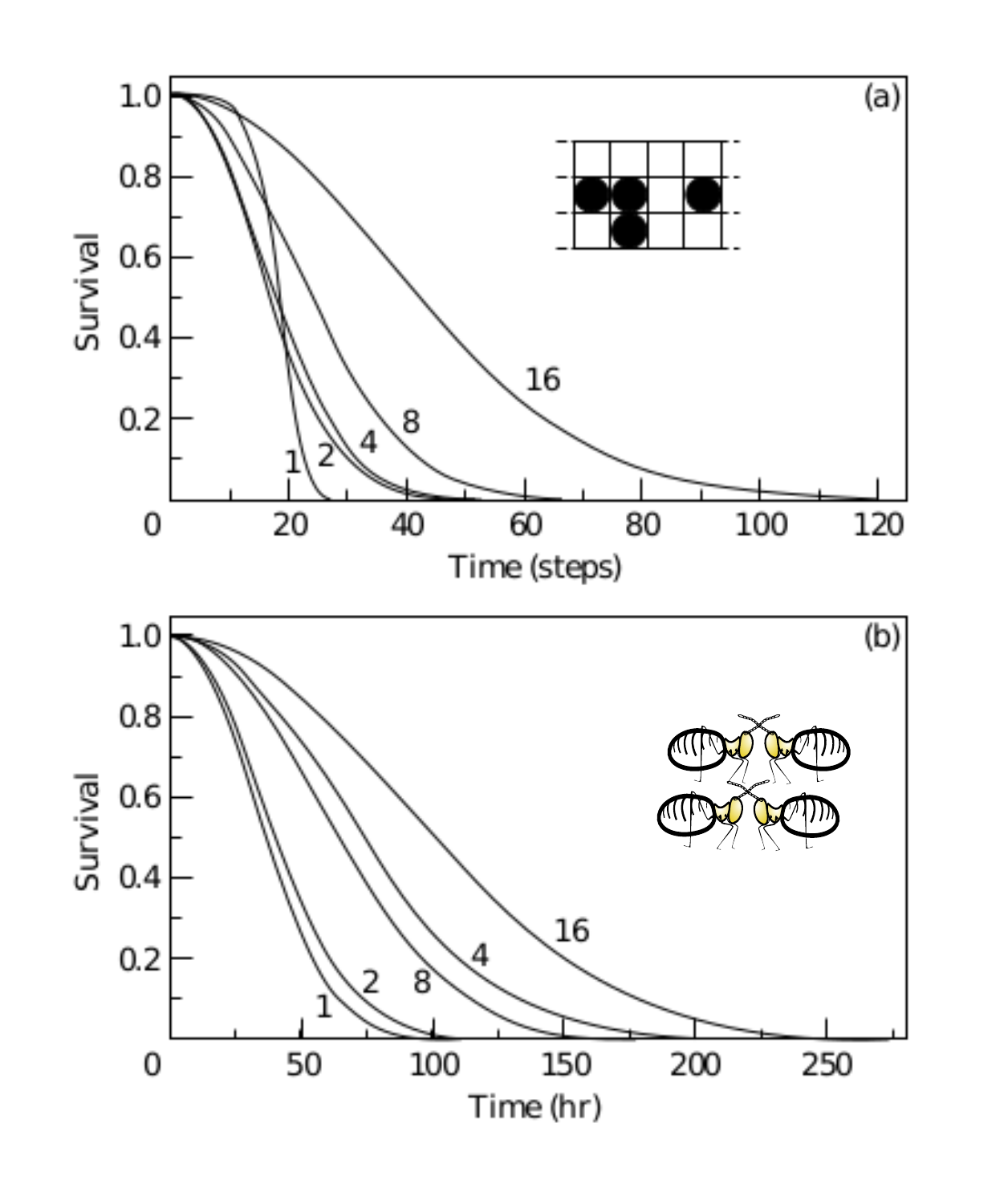}
  \caption{Qualitative comparison between survival of simulated and real termites using a Weibull distribution function. The same qualitative patterns are observed for simulated (a) and real termites
(b) suggesting that social interactions are important in the mechanism leading to longer survival in grouped termites. Modified from \cite{Miramontes.DeSouza.1996}. }
  \label{fig:survival-automata}
 \end{figure}
\end{center}

Evidence points that this is likely: an experiment has shown that survival of termites depends on group size \cite{DeSouza.etal.2001} (Fig.~\ref{fig:survival-grouped}). This survival is ``social facilitated'' (Fig.~\ref{fig:survival-automata}), closely resembling the ``survival'' of mobile cellular automata also modelled in groups of varying size and governed by two simple rules only: (i) individual energy decays with time and (ii)  when moving around, a automata can recover some energy by interacting to another \cite{Miramontes.DeSouza.1996}. 

It has been also showed that interindividual interactivity  at optimal density is key to this survival: under low density, termites are more mobile and survive longer and these experimental procedures reveal the presence of phase transitions in termite social behaviour \cite{DeSouza.Miramontes.2004,Miramontes.DeSouza.2008}.  

\begin{center}
 \begin{figure}
 \centering
  \includegraphics[scale=0.7]{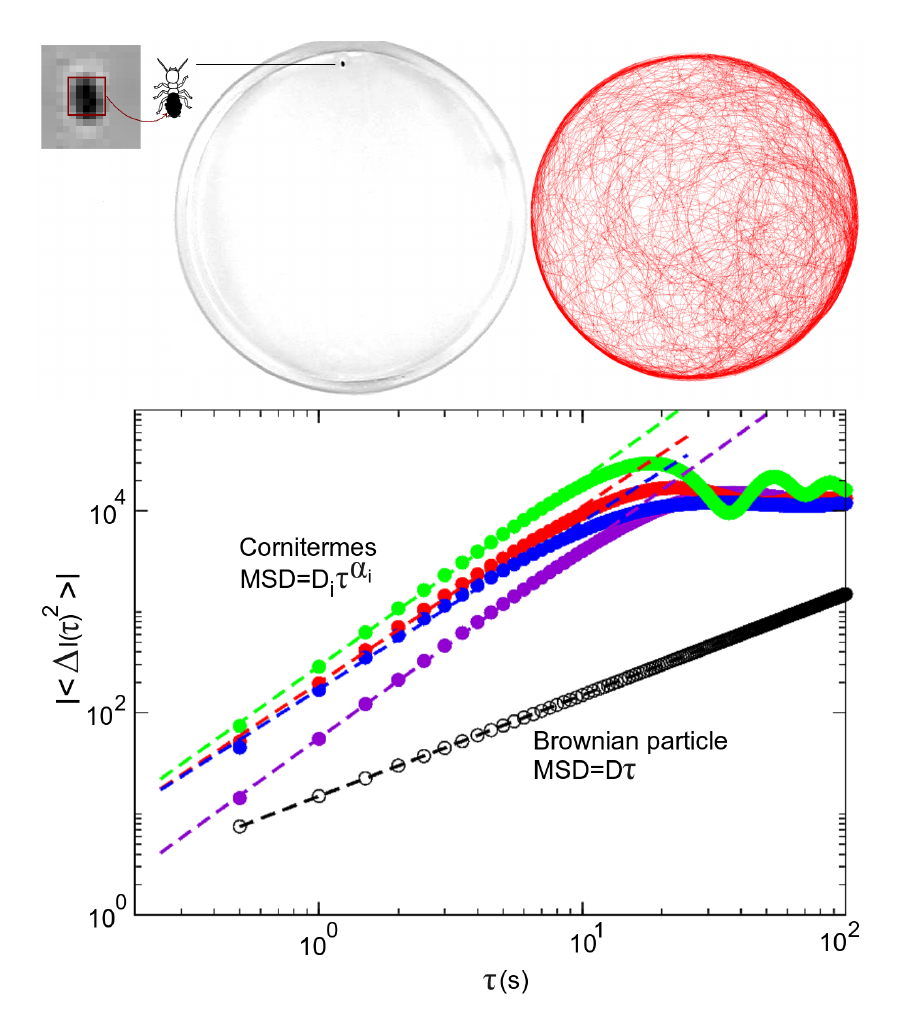}
  \caption{Upper row: video-recorded path of a \textit{Cornitermes cumulans} termite worker walking for c.a. 6 hours in a 205 mm Petri dish, sampled every 0.5 sec hence generating a trajectory of c.a. 35,000 points. The termite is seen as a 5$\times$5 pixel image of approximately 4.7 mm$^2$. Lower row: analysis of mean square displacement (MSD) of the trajectories of four termites, shot individually as in the above row, showing evidence for anomalous diffusion which is one of the signatures of Lévy walks. Other signatures for this walking pattern have been also spotted: see \cite{Miramontes.etal.2014} for more details. Anomalous difusion is characterized here by the fact that the mean squared displacement grows faster than it does in the normal diffusion of a Brownian particle whose curve, depicted as the lowest one, has a MSD scaling exponent $\alpha=1$. For the curves above this curve (each one representing a given termite in a given Petri dish), the scaling exponents are, respectively, $\alpha=\{1.90, 1.66, 1.75, 1.86\}$. Modified from \cite{Miramontes.etal.2014}.}
  \label{fig:termite-path-msd}
 \end{figure}
\end{center}

If mobility is key to interactivity and this latter is adaptive, one should expect mobility patterns of termites to follow an optimized search strategy. This is precisely what we observed in movement patterns of a individual termites: a Lévy-like route was established \cite{Miramontes.etal.2014}. We described exploratory spatial behaviour in isolated termite workers kept in large containers, free from the constrained movements they experience within tunnels (Fig.~\ref{fig:termite-path-msd}). In this way we were able to assess individual free exploratory behaviour in clueless environments and away from social interactions. We concluded that their searching patterns are compatible with scale-free strategies based on a fractal exploration of space and that these are key to the efficient flow of information between nestmates, thereby providing expressive hints on how self-organization underlies social cohesion.

In conclusion, simple repeated interactions between individuals can produce complex adaptive  patterns at the level of the group. Eusociality may, indeed, be an emergent phenomenon. 

\subsection{Interspecific interactions and termitophily}
Because interactivity and awareness-of-the-other lie in the very heart of organic evolution, a deeper insight tempting arises. The awareness of the other has been proposed to be one of the traits helping organisms to cross the barrier to sociality and eusociality \cite{Wilson.2012}.  In fact, one of the traits common to all transistions in evolution \cite{Maynard-Smith.Szathmary.1997.majortransitionsin} is that entities come about together to form other entities: molecules compose chromosomes which compose cells, and these compose multicellular organisms, which compose colonies. Moreover, many complex entities we see today are in fact a combination of distinct ones: our cells are in fact the product of the engulfing of once free-living prokaryotes by a primitive eukaryotic cell. This prokariotes are now what we know as ``mitochondria''\cite{Margulis.Sagan.1990.Origins}, which is now unable to live outside the the cell. A similar scenario applies to chloroplasts in plants. 

Maybe not surprisingly, termites provide us with a macroscopic example of such strict integration: in the nest of many termite species, other termite species and other invertebrates coexist \cite{Cristaldo.etal.2012,Marins.etal.2016.Termitecohabitationrelative} and this is called ``termitophily''. Moreover, some do it in an obligatory way: they can not survive outside such nests \cite{Florencio.etal.2013.Diet}. Understanding such symbiosis may help us to unveil the process by which termites deal with strangers and, ultimately, the process by which two distinct entities come about to form  single one. 

\begin{center}
 \begin{figure}[!t]
 \centering
  \includegraphics[scale=0.3]{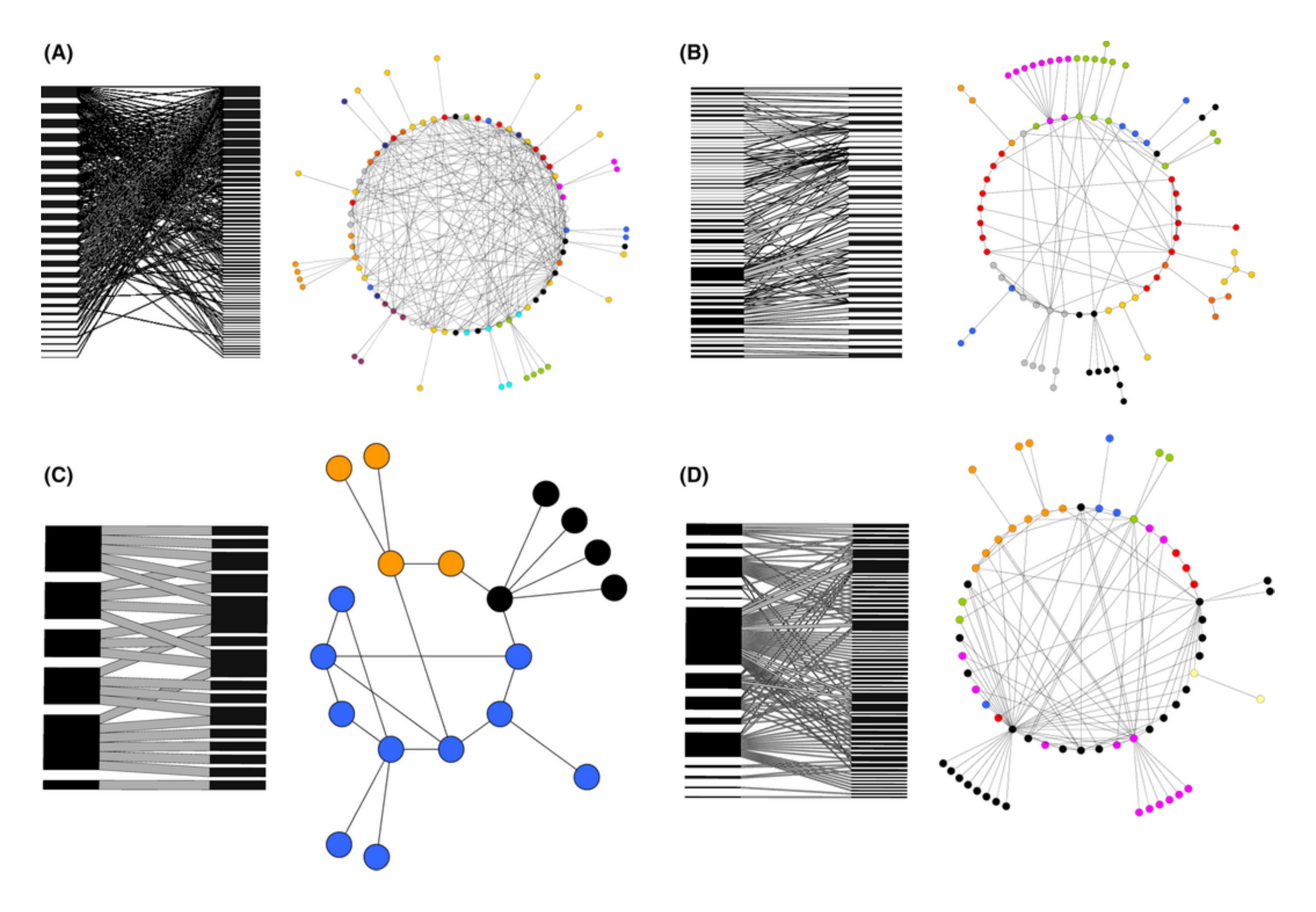}
  \caption{Visual comparison of (A) the plant–pollinator interaction network, (B) the host–parasitoid interaction network, (C) the Cameroon and (D) the Brazilian termite–termite networks. Each panel shows bipartite projections that emphasize nestedness (left) and the circular projections that show the compartmentalization structure (right), where colors indicate compartments. A comparison of the properties emerging from these networks has shown that termite-termite communities C \& D resemble more closely to mutualistic than to antagonistic communities depicted respectivelly in A \& B.   Modified from \cite{Campbell.etal.2016.Top}, where more details on the analytical procedure are given.}
  \label{fig:network-termitophile}
 \end{figure}
\end{center}

Again, computing models may be good tools to reveal the underlying forces. One of such approaches is the use of network analysis to reveal hidden termite-termite interactions. We contrasted the emergent network properties of multispecies termite communities cohabting the same termitarium with the properties of detail-rich reference communities with known modes of interaction \cite{Campbell.etal.2016.Top} (Fig.~\ref{fig:network-termitophile}). The studied termite networks overlapped more closely with mutualistic plant–pollinator communities than to antagonistic host–parasitoid communities. The analysis raised the hypothesis that termite–termite cohabitation networks may be overall mutualistic. More broadly, it provided support for the argument that cryptic communities may be analyzed via comparison to well-characterized communities.

Another approach is to model, through genetic algorithms and artificial life simulations, how do inquilines in termites manage to cohabit the nest along with the building species? Why do host colonies tolerate intruders? What kind of evolutionary pressures can explain the origin of this phenomenon that apparently benefits only the inquilines? These issues are difficult to investigate using classic methods at disposal of biologists. Computer simulation models can help to test and generate new hypothesis that can be subsequently verified with targeted experiment in nature. A recent collaboration between Federal University of Viçosa and Aberystwyth University has started aiming to develop studies using such an approach.

\section{Conclusion}
Symbiotic processes occurring at the intra- and inter-specific level within termitaria area key to understand basal mechanisms leading to the major transitions in evolution. Because such processes involve interindividual interactions and may result and emergent phenomena, their full understanding may be better achieved by the use of computing tools and models rather than classical biological methods of analysis.

\section*{Acknowledgments}
This is an excerpt of a talk given by ODS at the 14th International Conference on the Simulation of Adaptive Behavior, at Aberystwyth, Wales, on 23-26 Aug 2016. Since it is intended as a review, this text stems from previous works by the authors and, in particular, it draws heavily from \cite{Campbell.etal.2016.Top,DeSouza.etal.2001,Florencio.etal.2013.Diet,Miramontes.DeSouza.1996,Miramontes.DeSouza.2014.Social,Miramontes.etal.2014}. The work described in this paper has been supported by The Brazilian National Council for Research Development (CNPq), the Coordination for the Improvement of Higher Personnel (CAPES), Minas Gerais State Foundation for Research Support (FAPEMIG) and the Newton Fund-CONFAP-FAPEMIG initiative (APQ-0811/15). ODS holds a CNPq Fellowship (CNPq PQ 305736/2013-2) This is contribution \#65 from the Lab of Termitology at Federal University of Viçosa (\url{http://www.isoptera.ufv.br}).

\newcommand{\ra}{$\rightarrow$}
\begin{table}
\caption{The major transitions in evolution, as proposed by Maynard-Smith \& Szathmáry \cite{Maynard-Smith.Szathmary.1997.majortransitionsin}. Note that all transitions involve some type of ``come and stay together'' (so called symbiosis) between discrete entities and that termite colonies are among the last transitions (depicted in boldface). Also, termitophily, the living together of termites and other organisms, would be posed right before the rise of true colonies, in the transition from solitary to colonial organisms. }
\begin{center}
\begin{tabular}{lcp{7cm}}
\toprule
Replicating molecules & \ra & Populations of molecules\\
Independent replicators & \ra & Chromosomes\\
RNA as gene and enzyme & \ra & DNA genes, protein enzymes\\
Bacterial cells (prokariotes) & \ra & Cells with nuclei and organelles (eukaryotes)\\
Asexual clones & \ra & Sexual populaitons\\
Single-celled organisms & \ra & Animals, plants and fungi\\
\textbf{Solitary individuals} & \ra & \textbf{Colonies with non-reproductive castes (ants, bees, termites)}\\
Primate societies & \ra & Human societies (language)\\
\bottomrule
\end{tabular}
\end{center}
\label{tab:transitions}
\end{table}
\newpage

\end{document}